\documentclass{article}

\usepackage{arxiv}

\usepackage[utf8]{inputenc} % allow utf-8 input
\usepackage[T1]{fontenc}    % use 8-bit T1 fonts
\usepackage{hyperref}       % hyperlinks
\usepackage{url}            % simple URL typesetting
\usepackage{booktabs}       % professional-quality tables
\usepackage{amsfonts}       % blackboard math symbols
\usepackage{nicefrac}       % compact symbols for 1/2, etc.
\usepackage{microtype}      % microtypography
\usepackage{lipsum}
\usepackage{soul}
\usepackage{amsmath}
\usepackage{url}
\usepackage{multirow}
\usepackage{amssymb}
\usepackage{subfig}
\usepackage{graphicx}

\newcommand{\zvec}{\mathbf{z}}
\newcommand{\Thetavec}{\mathbf{\Theta}}
\newcommand{\trans}{\mathrm{T}}

\newcommand{\argmin}{\mathop{\rm argmin}\limits}
\usepackage{paralist}

\title{Predicting Evacuation Decisions using Representations of Individuals' Pre-Disaster Web Search Behavior}

\author{
  Takahiro Yabe \\
  Lyles School of Civil Engineering\\
  Purdue University, USA\\
  \texttt{tyabe@purdue.edu} \\
   \And
 Kota Tsubouchi \\
  Yahoo Japan Corporation\\
  Tokyo, Japan\\
  \texttt{ktsubouc@yahoo-corp.edu} \\
   \And
 Toru Shimizu \\
  Yahoo Japan Corporation\\
  Tokyo, Japan\\
  \texttt{tshimiz@yahoo-corp.edu} \\   
  \And
 Yoshihide Sekimoto \\
 Institute of Industrial Science\\
  University of Tokyo\\
  Tokyo, Japan\\
  \texttt{sekimoto@iis.u-tokyo.ac.jp} \\
  \And 
  Satish V. Ukkusuri\\
  Lyles School of Civil Engineering\\
  Purdue University\\
  Indiana, USA \\
  \texttt{sukkusur@purdue.edu} \\
}

\begin{document}
\maketitle

\begin{abstract}
Predicting the evacuation decisions of individuals before the disaster strikes is crucial for planning first response strategies. 
In addition to the studies on post-disaster analysis of evacuation behavior, there are various works that attempt to predict the evacuation decisions beforehand.
Most of these predictive methods, however, require real time location data for calibration, which are becoming much harder to obtain due to the rising privacy concerns. 
Meanwhile, web search queries of anonymous users have been collected by web companies. 
Although such data raise less privacy concerns, they have been under-utilized for various applications. 
In this study, we investigate whether web search data observed prior to the disaster can be used to predict the evacuation decisions. 
More specifically, we utilize a \textit{session-based query encoder} that learns the representations of each user's web search behavior prior to evacuation. 
Our proposed approach is empirically tested using web search data collected from users affected by a major flood in Japan. 
Results are validated using location data collected from mobile phones of the same set of users as ground truth. 
We show that evacuation decisions can be accurately predicted (84\%) using only the users' pre-disaster web search data as input. 
This study proposes an alternative method for evacuation prediction that does not require highly sensitive location data, which can assist local governments to prepare effective first response strategies. 
\end{abstract}

% keywords can be removed

\section{Introduction}
% 矢部
Severe disasters such as the Tohoku tsunami (2011) and Hurricanes Harvey, Irma and Maria (2017) have caused mass evacuation activities due to damage on urban infrastructure \cite{yun2015evacuation}.
% ,chiaro2017reconnissance,mimura2011damage}.
Speedy and effective response to natural disasters is of utmost importance to numerous cities around the world, due to the increasing frequency and intensity of large scale hazards \cite{unisdr2012disaster}. 
For efficient disaster response, it is crucial to predict evacuation activities before the disaster strikes.
Such information can be used by decision makers to prepare effective strategies, such as the optimal allocation of emergency supplies and an efficient spatial distribution of evacuation shelters.
% For efficient disaster response, it is crucial to capture information about the affected populations from various dimensions. 
% One important dimension is the mobility patterns of individuals after disasters. 
% Understanding the evacuation mobility patterns during crisis is crucial for planning relief efforts, including the optimal allocation of various emergency supplies and distribution of evacuation shelters. 
Conventionally, surveys and census data have been used as primary sources to analyze evacuation decisions after disasters \cite{sadri2018role,sadri2017modeling,mesa2012household}.
More recently, studies have utilized large scale location datasets for post-disaster analysis (e.g. mobile phone GPS \cite{yabe2016framework,song2013intelligent}, call detail records \cite{bagrow2011collective,lu2012predictability}, Twitter Geo-tags \cite{wang2014quantifying}). 
Such works have provided insights on evacuation behavior through spatio-temporally detailed analysis.
However, such analyses often lack predictive power on what will happen in future disasters.
% However, past events have shown that it is still extremely difficult to predict evacuation activities beforehand by only using knowledge from past disasters. 

\begin{figure*}[t]
\centerline{\includegraphics[width=0.9\textwidth]{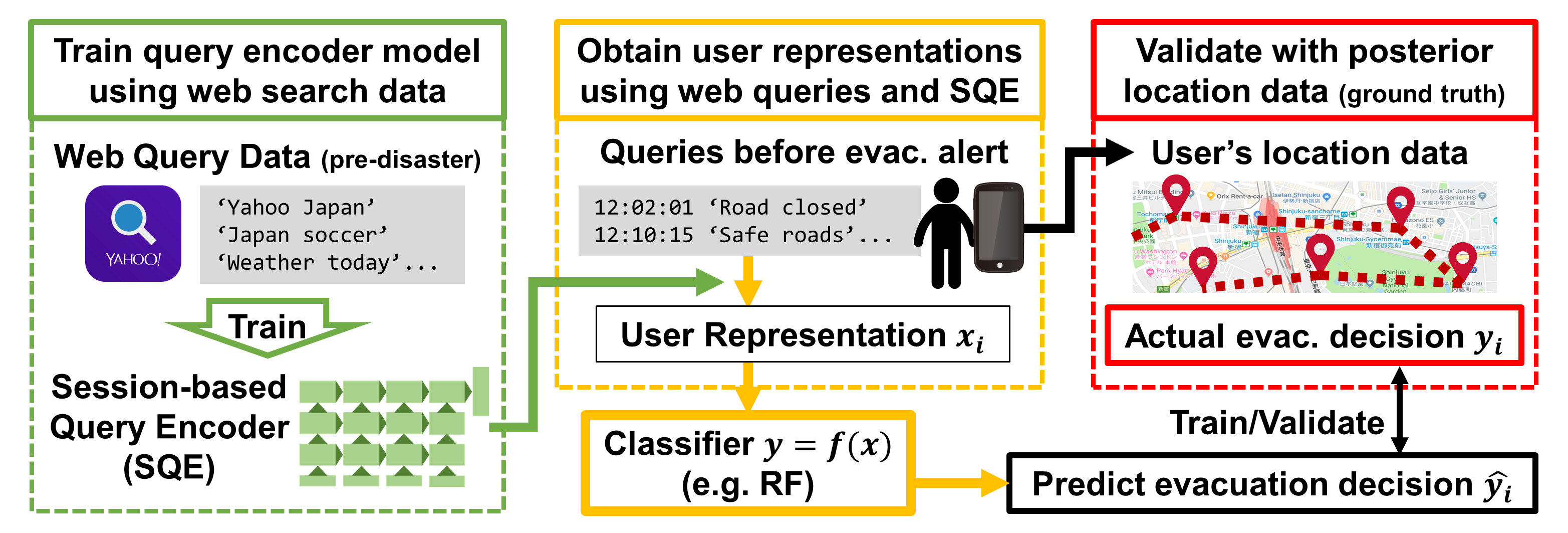}}
\caption{Overall framework of the study. Evacuation decisions are predicted using user representations generated from web search queries observed prior to the disaster, and are validated using location data as ground truth.}
\label{framework}
\end{figure*}

To overcome such drawback of post-disaster analyses, various studies have proposed online methods to predict near future evacuation mobility \cite{sudo2016particle,fan2015citymomentum,song2014prediction,song2013modeling,song2017prediction,jiang2018deepurbanmomentum}. 
Although experiments show the effectiveness of such methods, most of these predictive methods require real time location data for model calibration. 
Recently, it is becoming increasingly difficult to obtain and use real time location data due to rising privacy concerns \cite{de2013unique}. 
Thus, there is increasing demand for methods that can predict evacuation decisions using alternative data sources which raise less privacy concerns. 

% Another stream of research attempts to fill this gap by analyzing text data mainly posted on social media services by the affected individuals. 
% There exists an extensive literature on works that analyze Twitter posts during disasters to understand the sentiment dynamics, magnitude of the damage, and social interactions between users \cite{kryvasheyeu2015performance,ukkusuri2014use,kryvasheyeu2016rapid,sadri2018analysis}. 
% These works have greatly contributed to the understanding of the dynamic needs of affected individuals during disasters \cite{hyperlocal}, and has shown that social media data can be utilized for information retrieval in practice \cite{hughes2009twitter}. 
% However, there has been little efforts to utilize such text data to better understand evacuation mobility, mainly due to the unavailability of data that connects such text information with mobility traces.
% One candidate dataset for such investigation is the Twitter geo-tag dataset, however the data is temporally too sparse to analyze detailed evacuation behavior.
% Thus, despite the rising interest and activeness in research on both evacuation mobility analysis and disaster information retrieval using novel and large data sources, there is a disconnection between the two topics.

In this paper, we attempt to bridge this gap by proposing a method that utilizes web search queries collected from app users to predict evacuation decisions after disasters. 
Web search queries pose less threats to the users' privacy compared to location information, and are commonly collected by web service companies to improve browsing experiences. 
Despite this advantage, research on utilizing web search queries for various applications have been under-investigated. 
Konishi et al., which predicts future congestion in transit stations using transit app queries \cite{konishi2016cityprophet}, is the closest study to ours.
However, compared to transit queries where users explicitly search names of destination stations, web search queries have a much larger vocabulary set with latent meanings and subtle differences in expressions, making it much harder to utilize for mobility prediction. 
Thus, to overcome this difficulty, we utilize an LSTM model that learns the representations of queries by exploiting the web search behavior and the underlying search intent of users. 

Using web search data and location information (used only as ground truth for validation) of Yahoo Japan App users, we test the predictive performance of our method. 
Figure \ref{framework} shows the overall framework of our study. 
First, we train our session-based query encoders (\texttt{SQE}) using web search queries collected prior to the disaster. 
Then, we select users who are located within the disaster region, and learn their representations from query data observed prior to the disaster alert using the \texttt{SQE}.
We test the predictive performance of evacuation decisions using the user representations, by validating with actual evacuation decisions observed from location data. 
The intuition is that, users who decide to evacuate have common web search behavior traits before the disaster alert. 
Our approach overcomes the aforementioned drawbacks because it does not require location information of the users for prediction. 
We use real world data (both web search and location information) collected from users affected by the 2018 Japan Floods for experimentation. 

The main contributions of this paper are:
\begin{itemize}
    \item We present a novel approach that utilizes users' web search queries collected prior to the disaster alert to predict their evacuation decisions. 
    \item We clarify that evacuation decisions can be predicted with high accuracy using real world data collected during the 2018 Japan Floods. 
    \item We find that encoding the queries can improve the predictive accuracy of evacuation decisions, by capturing subtle differences in expressions and vocabulary.
\end{itemize}

The following sections are organized as follows. 
Section 2 provides details of the web search and location datasets used in this study. 
We introduce our methodology in Section 3, and present experimental validation results in Section 4. 
The results are discussed in Section 5, related works are introduced in Section 6, and conclusions are made in Section 7.

\section{Dataset}
% Yahoo Japan Corporation\footnote{https://about.yahoo.co.jp/info/en/} collects various data from their users to improve services. 
In this study, we utilized web search data and location information collected by Yahoo Japan Corporation\footnote{https://about.yahoo.co.jp/info/en/}. 
This is a unique dataset that contains both web search data and high granular location information of users, with the same set of IDs.
Thus, this dataset provides us a valuable opportunity to explore the predictability of mobility decisions using web search behavior. 

\subsection{Web Search Data}
\begin{figure}[t]
\centerline{\includegraphics[width=0.55\columnwidth]{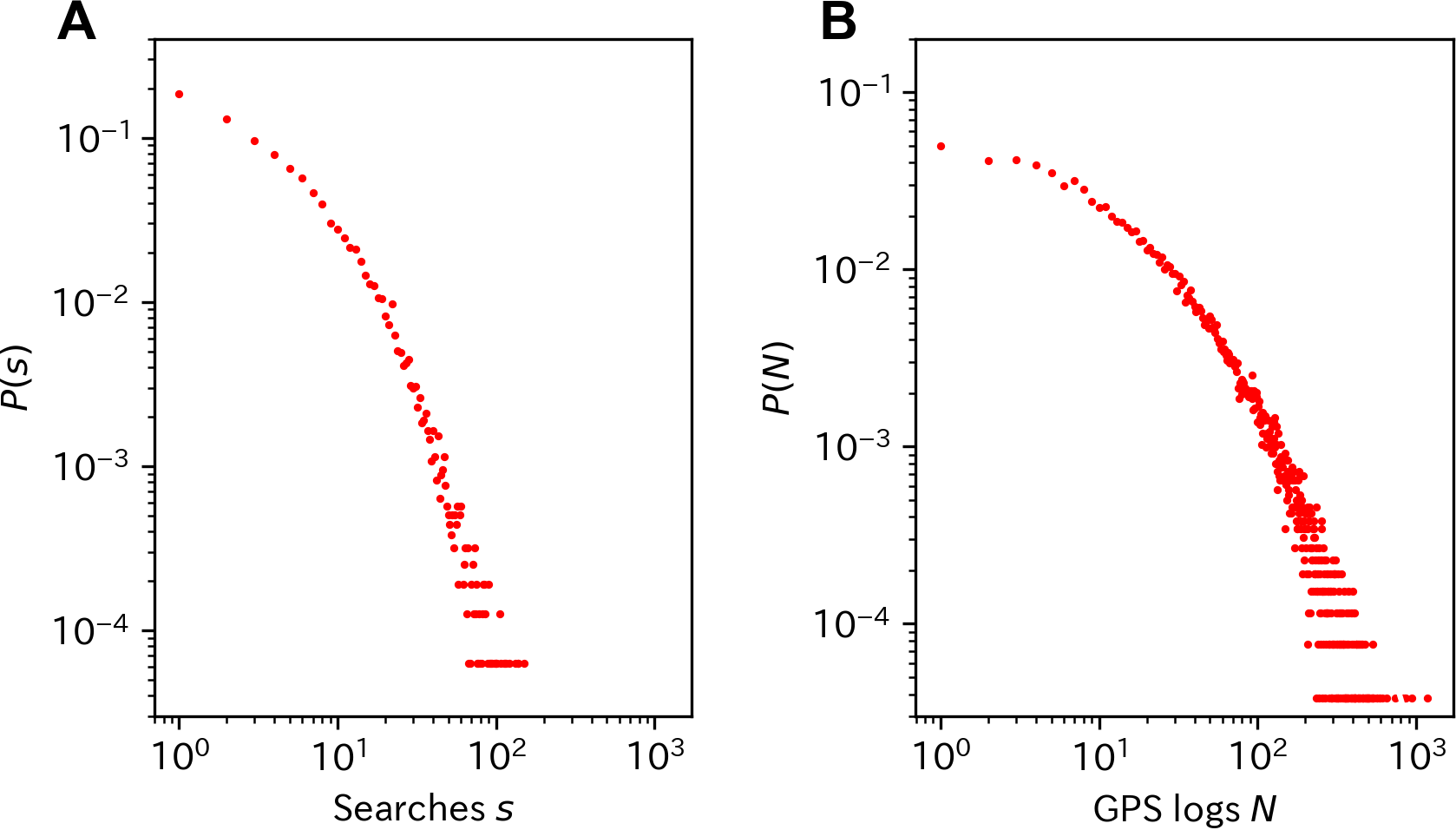}}
\caption{Probability distribution of: \textbf{A)} web search queries per individual on a given normal day \textbf{B)} GPS location data points per user on a given day and.}
\label{powerlaw}
\end{figure}

% 矢部
The ``Yahoo Japan app'' is a popular smartphone app in Japan, which is a platform that provides various services including web search, shopping, weather information and disaster alerts. 
% There are over 20 million active users in Japan, and 
Yahoo Japan collects the web search queries of the users to improve the web search quality. 
Out of all users, a fraction of the users hold a Yahoo Japan user ID, which allows users to personalize and customize their services.
We collected the web search data of users that have agreed to provide their data for research purposes. 
Panel A in Figure \ref{powerlaw} shows that the probability distribution of the number of searches per day follows a truncated power law, with mean of around 12. 
 
\begin{figure}[t]
\centering
\includegraphics[width=0.55\columnwidth]{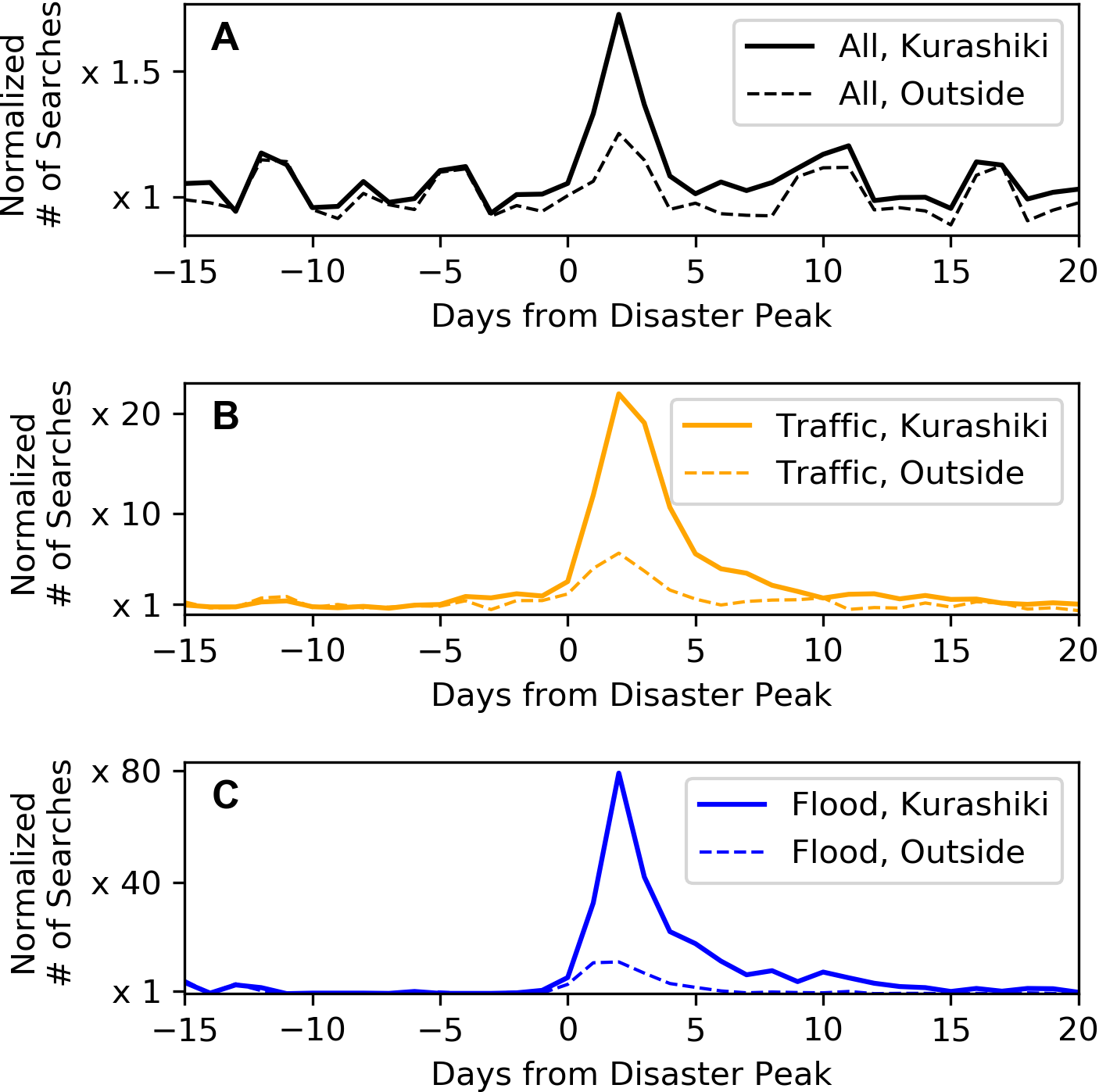}
\caption{Comparison of web search counts between affected and non-affected users before, during and after the disaster. \textbf{A)} Total searches. \textbf{B)} Traffic related searches. \textbf{C)} Flood related searches.} 
\label{searchcount}
\end{figure}

Figure \ref{searchcount} compares the frequencies of web searches performed by the users inside and outside the disaster affected area (Kurashiki), before, during and after the disaster date. 
Equal number of users were randomly chosen for both population groups (in and out of Kurashiki). 
% Thus, the absolute value of the number of searches directly reflect the search probabilities. 
Panel A shows that the total number of searches increase more for users located inside the disaster affected area compared to those who are not. 
% This result suggests that disaster affected users utilize web search to collect various information to prepare and respond to emergency situations. 
Panels B and C show that in addition to the total number of web searches, users directly affected by the disaster tend to search more about traffic information and flood information compared to users outside the disaster area. 
% Note that the number of such searches also increase for users outside the disaster affected area as well (despite the small magnitude), which reflects the rise in the interest in the disaster situations from the general public.
These empirical analysis of web search data clarifies our intuition that users' web search behavior are indeed affected by natural disasters, motivating us to use such data to predict evacuation decisions. 
In this study, as shown in Figure \ref{framework}, web query data collected prior to the disaster are used to train the session-based query encoders (\texttt{SQE}), and the users' web search queries prior to the evacuation alert are used as input of the \texttt{SQE} to generate user representations.
% Those representations will then be used as input to predict evacuation decisions.

\subsection{Mobile Phone Location Data}
% 矢部
In addition to web search data, the Yahoo Japan app also collects location information of users in order to send only relevant notifications to the users. 
The users in this study have accepted to provide their location information. 
The data are anonymized so that individuals cannot be specified, and personal information such as gender, age and occupation are unknown. 
% Each location observation point is stored as a GPS record in Yahoo Japan's internal server. 
Each GPS record consists of a user's unique ID (random character string), latitude, longitude, date and time. 
% The acquisition frequency of the GPS data changes according to the movement speed of the user. 
% If it is determined that the user is staying in a certain place for a long time, data is acquired at a relatively low frequency, and if it is determined that the user is moving, the data is acquired more frequently. 
% By reducing the number of times data is acquired using this algorithm, it is possible to reduce the burden on the user's smartphone. 
% On average, about 40 points are observed per day per user, . 
Panel B of Figure \ref{powerlaw} shows that the probability distribution of the number of observations per user per day is heavy tailed, with a mean of around 40 GPS points, allowing us to infer the major staying locations of each user. 
The GPS data collected by Yahoo Japan Corporation has a sample rate of about 2\% of the population, and past studies suggest that this sample rate is enough to grasp the macroscopic urban dynamics \cite{yabe2016framework,nishi2014hourly}. 
% Using such dense location data, we are able to accurately detect where each individual is staying overnight for each day before, during, and after the disaster.
However, privacy concerns on location information are recently increasing, and it is extremely difficult to collect and use such data in real time.   
Therefore in this study, we assume that real time location data cannot be obtained for prediction, and thus we use only the web query data to predict evacuation decisions.
The location information are used in this study as ground truth to validate our predictions.

\section{Methodology}
% \subsection{Problem Definition}
% \textcolor{green}{Yabe}

\subsection{Preliminaries}
\subsubsection*{Definition (\textbf{Web Search Session})} A user's web search behavior can be observed as a sequence of web search queries performed by the user.
Usually, such continuous sequence of searches within a short time period are performed under a consistent underlying search intent. 
We define these short sequences of searches as ``web search sessions'', and utilize the latent relationships between queries within the same session to encode the representations of each query. 
Section \ref{sqnprep} explains how we obtain web search sessions from the web search dataset.

\subsubsection*{Definition (\textbf{Session-based Encoding})} A user's web search session is governed by a consistent underlying search intent, which can be used to infer the semantics of each query. 
For example, from a user's search session with 4 queries: \textit{``New York'' $\rightarrow$ ``New York Metro Station'' $\rightarrow$ ``New York sightseeing'' $\rightarrow$ ``Times Square''}, we can infer that \textit{New York} is a city (possibly a sightseeing city), and that \textit{Times Square} is one of the sightseeing spots in \textit{New York}, without having prior knowledge on these individual queries. 
The idea of \textit{Session-based Encoding} is that by modeling a vast number of search sessions, we are able to extract the latent semantics of each query without having any prior knowledge about them. 
Sections \ref{sqnmodel} and \ref{sqnmodel2} introduce an RNN based method called \textit{Session-based Query Encoding} (\texttt{SQE}) that generates representations of queries $x\in \mathbb{R}^d$ using web search sessions. 

\subsubsection*{Definition (\textbf{Evacuation Decision})} During the disaster, each user decides whether or not to evacuate from his/her home location to a shelter or other locations. 
We define an evacuation decision as a binary variable ($y=1$ if evacuated, $y=0$ otherwise).
The ground truth of this binary variable is estimated using anomaly detection on location information, explained in Section \ref{ad}.

\subsubsection*{Problem Definition (\textbf{Evacuation Decision Prediction})} Our task is to, for each user $i$, predict evacuation decision $y_i$ based on his/her user representation $x_i$ generated by the \textit{Session-based Query Encoding} (\texttt{SQE}) model using web search queries observed prior to the disaster alert. 
We test multiple methods to generate $x_i$ using the user's web search query data, in Section \ref{exp}.

% \subsection{Web Search Behavior Representation} 
\subsection{Session-based Query Encoding} 
\label{sqn}
% 清水さん　手法について1ページ程度でお願いいたします。図とかもあるといいと思います。
% \textcolor{blue}{Shimizu-san}

We design our model so that it can produce query representations that reflect the search intent of the users within a search session.
The models are trained by predicting the next query given a query in the search session.
This way, the models can learn representations via the consistent underlying search intent of different queries, since the search intent usually stays the same within a session.
We name this type of models as ``session-based query encoders'' (\texttt{SQE}) due to this key characteristic.
Further, we build 2 types of \texttt{SQE} for different types of inputs; an \texttt{SQE} for an input with a single query (``\textit{Session-based Single Query Encoder} (\texttt{SSQE})'') that has an LSTM RNN structure, and an \texttt{SQE} for an input with multiple queries (``\textit{Session-based Multiple Query Encoder} (\texttt{SMQE})''), that has a hierarchical LSTM RNN structure.

% In this work, we need models to produce queries' representation vectors well reflecting the search intent.
% To obtain such models, we use two types of RNNs; one is an LSTM RNN which produces a vector for an input query, and the other is a hierarchical LSTM RNN which produces a vector for multiple input queries.
% The models are trained by predicting the next query of a given query in the search session.
% As the search intent usually does not change within a session, in this way, the model becomes able to recognize queries with different surface representations but involving the same search intent as such.
% In other words, in this way, the model becomes able to understand the search intent of a given query.

\subsubsection{Dataset Preparation} \label{sqnprep}
First, we extracted web search records from Yahoo Japan's web search service, from randomly determined 75 three-hour blocks between January 2015 and July 2018. 
% to prevent sparse sessions.
Each record has three attributes: a user's unique ID, query text, and the time-stamp.
Queries searched by the same user were then grouped into sessions, using two minutes as a threshold for the session timeout.
% While the standard threshold is 30 minutes, we shortened it by large to make the search intent within each session more consistent. 
While the standard threshold is 30 minutes, we used a significantly shorter threshold to ensure that the search intent within each session is consistent. 
% why not 30 minutes??? 
Sessions with only one query were discarded, as they are not applicable for next query prediction. 
Also, sessions with more than ten queries were truncated after the tenth query to prevent learning from exceptionally long sessions.
As a result, we obtained 301M sessions containing 804M queries.
We put 10,000 sessions aside for validation, 20,000 for testing and the rest for training.
We call this the ``\textit{query session dataset}''. 
We also constructed the ``\textit{query pair dataset}'' by extracting consecutive query pairs from sessions.
This dataset contains 503M, 16,694, and 33,166 pairs in the training, validation, and test set, respectively.

% \subsubsection{Dataset Preparation} \label{sqnprep}
% First of all, we extracted search log records of Yahoo Japan's Web search service. 
% The one record has three attributes: a user's unique ID, query text, and the time-stamp.
% Then, we grouped them into sessions, gathering queries inputted by a user, using two minutes for the session timeout.
% Sessions with only one query were discarded here as they are not applicable to next query prediction. 
% Also, for sessions with more than ten queries, we truncate them after the tenth query so as to prevent learning too much from exceptionally long sessions.
% The duration of the log from which we sampled the records was Jan. 2015 to Jul. 2018.
% If we had sampled records from the log of this duration completely randomly, they would have become sparse and made it difficult to recognize sessions.
% To avoid this, actually we first randomly determined three-hour blocks, 75 of them in this work, in the duration and extracted log records from each block.
% In this manner, we obtained 301M sessions containing 804M queries.
% From this set of sessions, we prepared two datasets.
% One is the sessions themselves, and we call this the query session dataset. 
% We put 10,000 sessions aside for validation and 20,000 for testing and used the rest for training.
% The other dataset is the query pair dataset which we prepared by extracting consecutive query pairs from sessions.
% The numbers of pairs in the training, validation, and test set are 503M, 16,694, and 33,166 respectively.

\subsubsection{Session-based Single Query Encoder (\texttt{SSQE})} 
\label{sqnmodel}
We used a character-based 3-layer LSTM RNN \cite{hochreiter97:_long_short_term_memor} as the main component of our \texttt{SSQE} model to encode a query, implementing it with LSTM formulation given by Graves \cite{graves13:_gener_sequen_with_recur_neural_networ}.
The size of the character embedding is 256, the hidden layer size of LSTM is 1024, and a fully-connected layer producing the final 128-dimension query representation is attached to the last timestep of the top-layer of the LSTM block.
The size of the character vocabulary is 6000, which is large enough to include all the Japanese characters for daily use defined in the government guideline.
In total, the model has 26M parameters.
We trained it using the \textit{query pair dataset}, using cosine similarity as a cost function so that representations of two queries become closer if they belong to an existing pair.
More specifically, for a query $Q$ and its next query $D$ in the session, the cosine similarity $R_{\Thetavec} (Q, D)$ between queries $Q$ and $D$ is defined as: 
\begin{align}
  R_{\Thetavec} (Q, D) = \frac{\zvec_Q^{\trans}\zvec_{D}}{\lVert \zvec_Q \rVert \lVert \zvec_{D} \rVert},
\end{align}
where $\Thetavec$ denotes model parameters in the encoder which generates 128-dimensional representations $\zvec_Q$ and $\zvec_D$ for the two queries $Q$ and $D$, respectively \cite{huang13:_learn_deep_struc_seman_model,palangi14:_seman_model_with_long_short}.

To learn representations so that the representations of a query pair comes close to each other, we consider probability $P_{\Thetavec}(D | Q)$ for query pair $Q$ and $D$, where there are five choices $\{ D^1, \ldots,$ $D^5 \}$:
\begin{align}
  P_{\Thetavec} (D_i^k | Q_i) = \frac{\exp(\beta R_{\Thetavec}(Q_i, D_i^k))}{\sum_{j=1}^5 \exp(\beta R_{\Thetavec}(Q_i, D_i^j))},
\end{align}
which is obtained by feeding values of $R_{\Thetavec}(Q_i, D_i^k)$ with $k = 1, \ldots, 5$ into the softmax function.
The index $i$ denotes that the pair $(Q_i, D_i^1)$ is the $i$th record in a dataset.
We set the correct query pair as $k=1$ ($D_i^1$), and randomly picked negative samples $\{ D_i^2, D_i^3, D_i^4, D_i^5 \}$ during training.
The inverse temperature coefficient $\beta$ was set to 10 to make the cross entropy loss large enough.
The correct query choice among the five $D^k$ is $k = 1$, so the cross entropy loss $l$ for the $i$th query pair can be written as
\begin{align}
  l_{\Thetavec}(Q_i, D_i^1) = -\log P_{\Thetavec}(D_i^1 | Q_i).
\end{align}
Then, we find the optimal parameters $\hat{\Thetavec}$ by minimizing the total loss $L$ for all the query pairs in the dataset:
\begin{align}
  \hat{\Thetavec} 
    &= \argmin_{\Thetavec} L_{\Thetavec} = \argmin_{\Thetavec} \sum_i l_{\Thetavec}(Q_i, D_i^1).
\end{align}

\subsubsection{Session-based Multiple Query Encoder (\texttt{SMQE})} 
\label{sqnmodel2}
\begin{figure}[t]
\centering
\includegraphics[width=0.5\columnwidth]{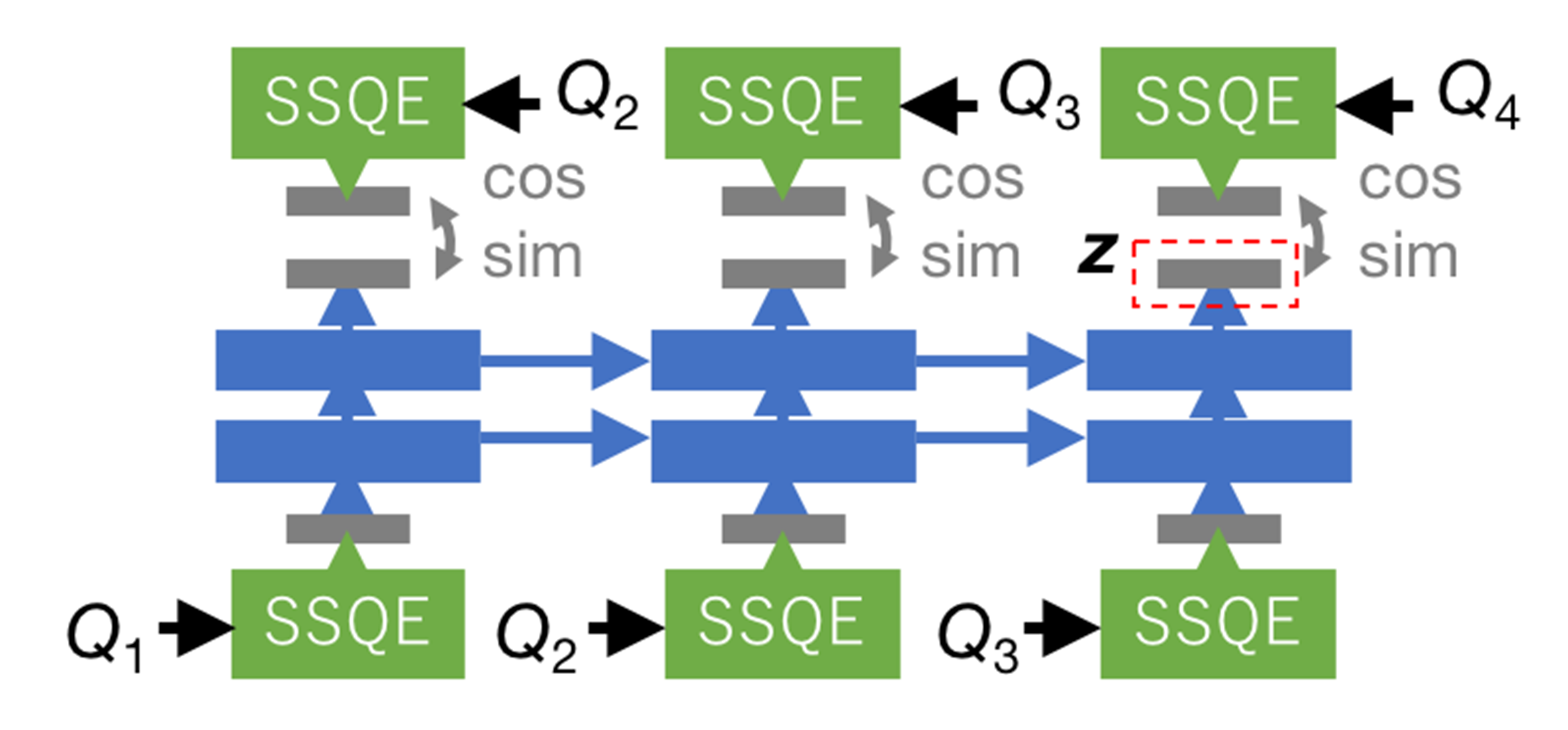}
\caption{\texttt{SMQE} processing a four-query session in the training phase. The model outputs the representation $\zvec$ of the query session.}
\label{smqe_arch}
\end{figure}
% 図、とてもいいと思います！スペース的な話になりますが、縦の長さをもう少しコンパクトにできますか？あと、最終的なrepresentationは赤で囲まれたものだと思うんですが、それを強調したいですね
% 圧縮版、作成してみました。赤は書かれている通り、inference時はこれを使います、というもので 本文でその旨補足したいところです
% ありがとうございます！とてもGOODです！
% 説明も素案だけ入れてしまいます。また適宜修正入れてくださいー → とりいそぎ該当セクションの最後に書き足しました
% ありがとうございます！あと、Figure 5 についても、バーグラフのところを2/3くらいの高さにすることってできますか？あとは、related worksあたりを少し削ろうかと思っています。
% あ、ちょうど今やっていたところでした。cos_sim_words2.pngを追加。・・・ちょっとフォントが小さすぎますのでもう少し試行錯誤します。例を減らした版（椎名林檎抜き）も含め。
% 個人的には椎名林檎の例は面白いので、なんか残したいです(笑)　バーを少し細くすればフォントすこし大きくできないですかね。。。？
% バーの横幅圧縮はなぜか出来ず、、一旦バーグラフの高さを抑えた版にしてみました。前よりは低くなりましたがまだ嵩張りますね。。
% ちょっと作ってみます！
% こちらも縦幅圧縮のコツが分かった気がします。mathematicaなんですが、aspect ratioの指定とマウス操作でそれらしい形に
% 本当ですか！ありがとうございます！！

To encode multiple queries at once, we used a hierarchical architecture \cite{Sordoni:2015:HRE:2806416.2806493} in which the \texttt{SSQE} and another 2-layer LSTM RNN were combined to build the \texttt{SMQE}.
The hidden layer size of the 2-layer LSTM is 1024, and it is also equipped with a fully-connected layer to produce a 128-dimension representation ($\zvec$) of a given query sequence.
Figure \ref{smqe_arch} illustrates the \texttt{SMQE} processing a four-query session $\{Q_1, Q_2, Q_3, Q_4\}$ in the training phase.
In the inference phase, the vector $\zvec$ obtained at the final time step is used as the representation of the three queries $\{Q_1, Q_2, Q_3\}$.
In total, the \texttt{SMQE} model has 47M parameters.

The \textit{query session dataset} is used for training this model.
In the training phase, when a sequence of queries is given to the model, the \texttt{SSQE} reads each query, and the 1024-dimension hidden representation at the last time step of the \texttt{SSQE}'s top layer of each query is fed to the 2-layer LSTM RNN one by one.
The combined model produces a 128-dimensional vector ($\zvec$) at each time step of the query sequence.
The generated representations, which capture the context, are compared with those generated by the \texttt{SSQE} for the next queries, and the closeness of each representation are evaluated by using cosine similarity.
% , in the same manner as \texttt{SSQE}'s training.
% We call this resultant model the session encoder.
In a nutshell, similar to \texttt{SSQE}, \texttt{SMQE} learns representations by next query prediction, but unlike \texttt{SSQE}, it can capture the context of the current query session.
This training process makes it possible for \texttt{SMQE} to generate a contextual representation.

\subsubsection{Training}
\texttt{SSQE} was trained using Adam \cite{kingma14:_adam} for 700M iterations with a batch size 96, over the \textit{query pair dataset}.
% The training task was defined as a classification, where the model has to select the correct next query from five choices (that includes four negative samples), and 
The accuracy for the aforementioned 5-class classification using the validation set was 93.4\%.
Using \texttt{SSQE} with the best validation performance as the first stage of the hierarchical model, we trained \texttt{SMQE} using Adam for 500M iterations with the batch size 64, over the \textit{query session dataset}.
In this setting, the classification accuracy reached 94.1\%, which is higher than that of \texttt{SSQE}.
% The difference between 93.4\% and 94.1\% may look insignificant, but 
To the best of our knowledge, none of the previous models trained under non-contextual settings have exceeded 0.94 in accuracy for the next query prediction task.
% even when we used a larger model or kept training a model longer.

% \subsubsection{Training}
% We trained the query encoder using Adam \cite{kingma14:_adam} for 700M iterations with a batch size 96, over the query pair dataset.
% The task the model tackles in the training can be seen as classification, selecting the actual next query from five choices including four negative samples, and its accuracy for the validation set reached 93.4\%.

% Using the query encoder with the best validation performance as the first stage of the hierarchical model, we trained the session encoder using Adam for 500M iterations with the batch size 64, over the query session dataset.
% In this contextual setting, the classification accuracy reached 94.1\%, which is higher than that of the query encoder model.
% The difference between 93.4\% and 94.1\% may look insignificant, but we have never seen a model trained under non-contextual settings reaching accuracy exceeding 0.94 for this task even when we used a larger model or kept training a model longer.

\begin{figure}[t]
\centering
\includegraphics[width=0.5\columnwidth]{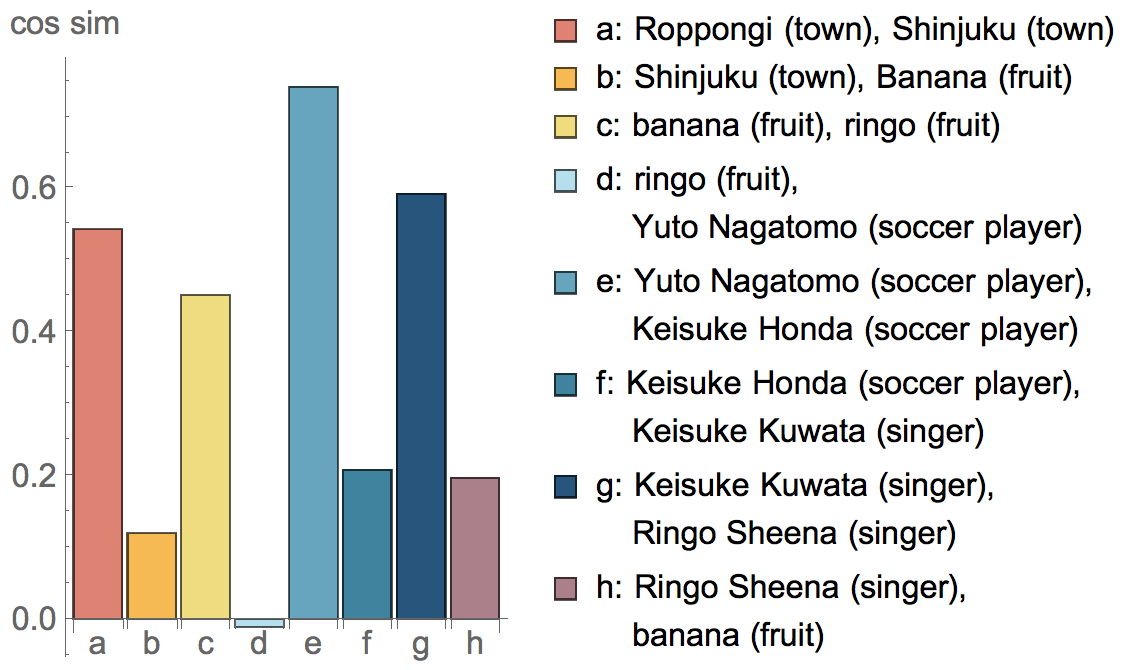}
\caption{Cosine similarity values for various query pairs.} 
\label{cos_sim_words}
\end{figure}
% 完璧です！本当にありがとうございます！！

% \begin{figure}[t]
% \centering
% \includegraphics[width=0.95\columnwidth]{cos_sim_groups.png}
% \caption{Averaged cosine similarity values for three word groups: randomly sampled queries, person names, and actor names.} 
% \label{cos_sim_groups}
% \end{figure}

\subsubsection{Examples of Similarity Between Query Words}
% We can confirm that the obtained query encoder model naturally groups queries belonging to the same category together.
Figure \ref{cos_sim_words} plots the cosine similarity values between various pairs of query representations. 
We can confirm that the similarity values are relatively high for query words from the same category (a,c,e,g), but are relatively low for those from different categories (b,d,f).
\texttt{SSQE} is able to discriminate similar but semantically different word pairs as well. 
Bar plot (h) shows that although the singer's name ``Ringo Sheena'' includes a fruit ``ringo (= apple)'' and is potentially confusing, the model is able to correctly distinguish between the singer and a fruit.

\subsection{Evacuation Detection using Location Data}
% 矢部
To validate the performance of our predictive model, we used location information collected from the users affected by the disaster. 
In reality, as explained in the introduction, location information are becoming more difficult to utilize in real time settings due to privacy concerns. 
In this study, location information were collected and used only in the validation phase, but not for predicting evacuation.

\begin{figure}
\centering
\includegraphics[width=0.5\columnwidth]{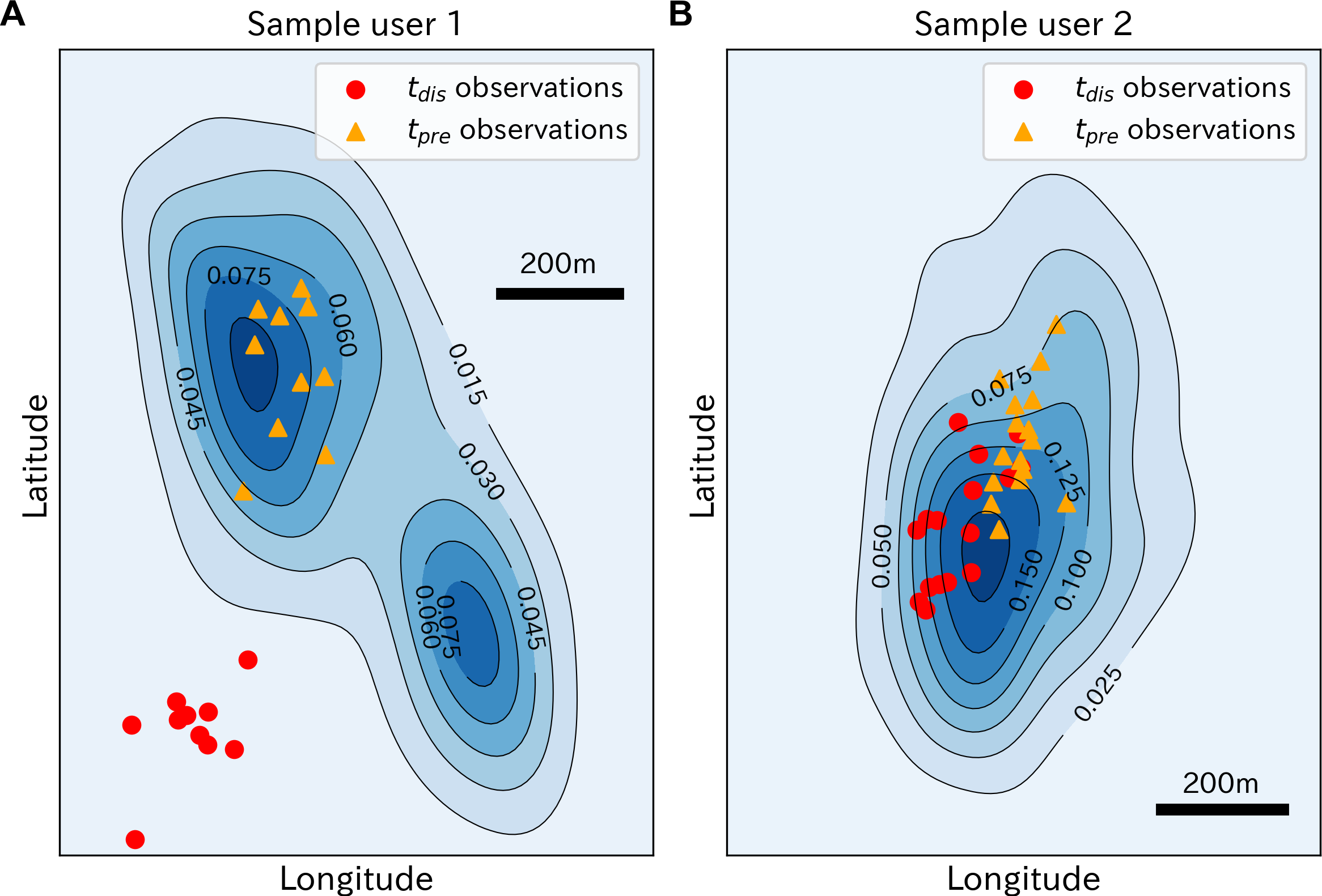}
\caption{Mobility anomaly detection outputs of 2 sample users. 
$\Hat{p}_i(\textbf{x}|d,t)$ (blue contours), observations before disaster (orange plots) and during disaster (red plots) are shown. 
Longitude and latitude values are hidden to protect users' privacy. 
\textbf{A)} User 1 was observed at unusual locations during the disaster, thus would receive a high anomaly score $\theta$ and label $y=1$ (indicating evacuation). 
\textbf{B)} User 2 was observed at usual locations during the disaster, thus would receive a low anomaly score $\theta$ and label $y=0$ (indicating no evacuation).} 
\label{contour}
\end{figure}

\subsubsection{Anomaly Detection} \label{ad}
This section explains the anomaly detection methods used to assign the label indicating evacuation or not ($y_i = \{1,0\}$) for each user using their location information. 
In a nutshell, the anomaly detection method assigns an anomaly score to each user based on the deviation of the users' post-disaster mobility patterns from his/her usual (pre-diaster) mobility patterns. 
Let us denote the learning period as $t_{learn}\in [t_0,t_l)$, pre-disaster period as $t_{pre}\in [t_l, t_d)$, and disaster period as $t_{dis} \in [t_d,t_e]$. 
$t_d$ is the timing when the evacuation alert was sent out. 
We estimated the spatial probability density of each user using data observed in $t_{learn}$, and then we compared anomaly scores of observations in $t_{pre}$ and $t_{dis}$ to assign labels $y$. 
If the anomaly score was higher during $t_{dis}$ than $t_{pre}$, this would indicate that the individual moved irregularly during the disaster (i.e. evacuation). 

Let us denote the set of observed location data of user $i$ as $O^i = \{o^i_1, o^i_2, \cdots , o^i_N \}$. 
Each observation is a 4-tuple of $o^i_n=\{i,t_n, x_n, y_n \}$ where $t_n$, $x_n$ and $y_n$ are the timestamp, longitude and latitude of the $n$-th observation of user $i$, respectively. 
Using location data observed during $t_{learn}$, we constructed the spatial probability density ($\hat{p}_i(\textbf{x}|d,h)$) of user $i$ conditional on the day of week and time of day. 
% Extending the idea of Cho et al., 
We assumed that each observation has a 2-D Gaussian probability density \cite{cho2011friendship}.
We estimated the spatio probability density $\hat{p}_i(\textbf{x}|d,h)$ of an observation $\textbf{x}=(x,y,t)$ of user $i$ conditional on day of week $d$ and time of day $h$ (hour) using multivariate Kernel density estimation with a Gaussian kernel, $K \sim \mathcal{N}(0,\Sigma)$: 
% which can be calculated by the following equation \ref{kde}.
\begin{equation}
    \hat{p}_i(\textbf{x}|d,h) = \frac{1}{N(J) (2\pi)^\frac{3}{2} |\Sigma|^{\frac{1}{2}}}  \sum_{\substack{j \in J}}
    \exp \Big\{ - \frac{1}{2} (\textbf{x} - \textbf{x}_j)^T \Sigma^{-1} (\textbf{x} - \textbf{x}_j) \Big\}
    \label{kde}
\end{equation}
where $J = \big\{ j \; | \; t_j \in t_{learn} , \; d(t_j)=d , \; h(t_j)=h  \big\}$ is a set of tags of observations used for density estimation, $N(J)$ is the number of tags in set $J$, $\Sigma = diag[\sigma^2_x, \sigma^2_y, \sigma^2_t]$, and $d(t)$ and $h(t)$ denote the day of week and hour of timestamp $t$, respectively. 
Then, for each user, the mean $p^i_{pre}$ and variance $(s^i_{pre})^2$ of probability values during the pre-disaster period and the mean probability value during the disaster period $p^i_{dis}$ are: 
\begin{eqnarray}
    p^i_{pre} = \mathbb{E}[\hat{p}_i(\textbf{x}|d,h)| \; t \in t_{pre}] \\
    (s^i_{pre})^2 = Var[\hat{p}_i(\textbf{x}|d,h)| \; t \in t_{pre}] \\
    p^i_{dis} = \mathbb{E}[\hat{p}_i(\textbf{x}|d,h)| \; t \in t_{dis}]
\end{eqnarray}
The anomaly score $\theta_i$ of individual $i$ was calculated by $\theta_i = \frac{p^i_{pre}- p^i_{dis}}{s^i_{pre}}$. 
Finally, we labeled users with high anomaly scores $\theta_i > \Tilde{\theta}$ as $y_i = 1$ (evacuated), and users with low anomaly scores $|\theta_i| < \Tilde{\theta}_l$ as $y_i = 0$ (not evacuated). 
$\Tilde{\theta}$ and $\Tilde{\theta}_l$ are threshold parameters used for labelling users.
All parameter values are specified in Section \ref{setup}. 

\subsubsection{Example Outputs}
Figure \ref{contour} illustrates how the anomaly detection method works using location data of 2 sample users. 
In each panel, the contour plots show the estimated spatial probability density $\hat{p}_i(\textbf{x}|d,h)$, orange points indicate observations before the disaster ($t_{pre}$), and red points indicate the observations during the disaster ($t_{dis}$).
The longitude and latitude values are hidden to protect the users' privacy. 
The values of $\Sigma$ that were used for the multivariate kernel density estimation were $\sigma_x = \sigma_y = 100 \text{ (m)}$ and $\sigma_t=1 \text{ (hour)}$. 
% Mean and standard deviation of the probability scores of observation points are annotated in the figure. 
% The scores are ... 
We can clearly observe that sample user 1 was observed before the disaster in his usual location. 
However, he was observed at very unlikely locations during the disaster, implying that this user evacuated to a safer location (e.g. evacuation shelter). 
Thus, this user will receive a high anomaly score $\theta$. 
In contrast, we can observe that sample user 2 was observed near the usual location even during the disaster, implying that this user did not evacuate. 
Thus, this user will receive a low anomaly score $\theta$. 

\section{Experimental Validation} \label{exp}

\subsection{Experiment Setup} \label{setup}
We tested our proposed method using data collected from the 2018 Japan Floods, which resulted in widespread devastating floods and mudflows across Japan from mid-June to mid-July in 2018. 
More than 220 people were confirmed dead across 15 prefectures, and more than 8 million people were advised or urged to evacuate across 23 prefectures\footnote{\url{https://www.cnn.com/2018/07/10/asia/japan-floods-intl/index.html}}.
In particular, Okayama and Hiroshima prefectures experienced severe flooding.  
% In the Mabi area of Kurashiki City, 51 people lost their lives due to flooding, and more than 1,200 hectares were flooded. 
Evacuation preparation notifications were sent out on July 6th 10PM, and the evacuation alert was issued on July 7th 1:30AM. 
The riverbank was overwhelmed by water 5 hours after the evacuation alert, at 6:52AM of July 7th. 

\subsubsection{Parameter Settings}
\begin{figure}
\centering
\includegraphics[width=0.5\columnwidth]{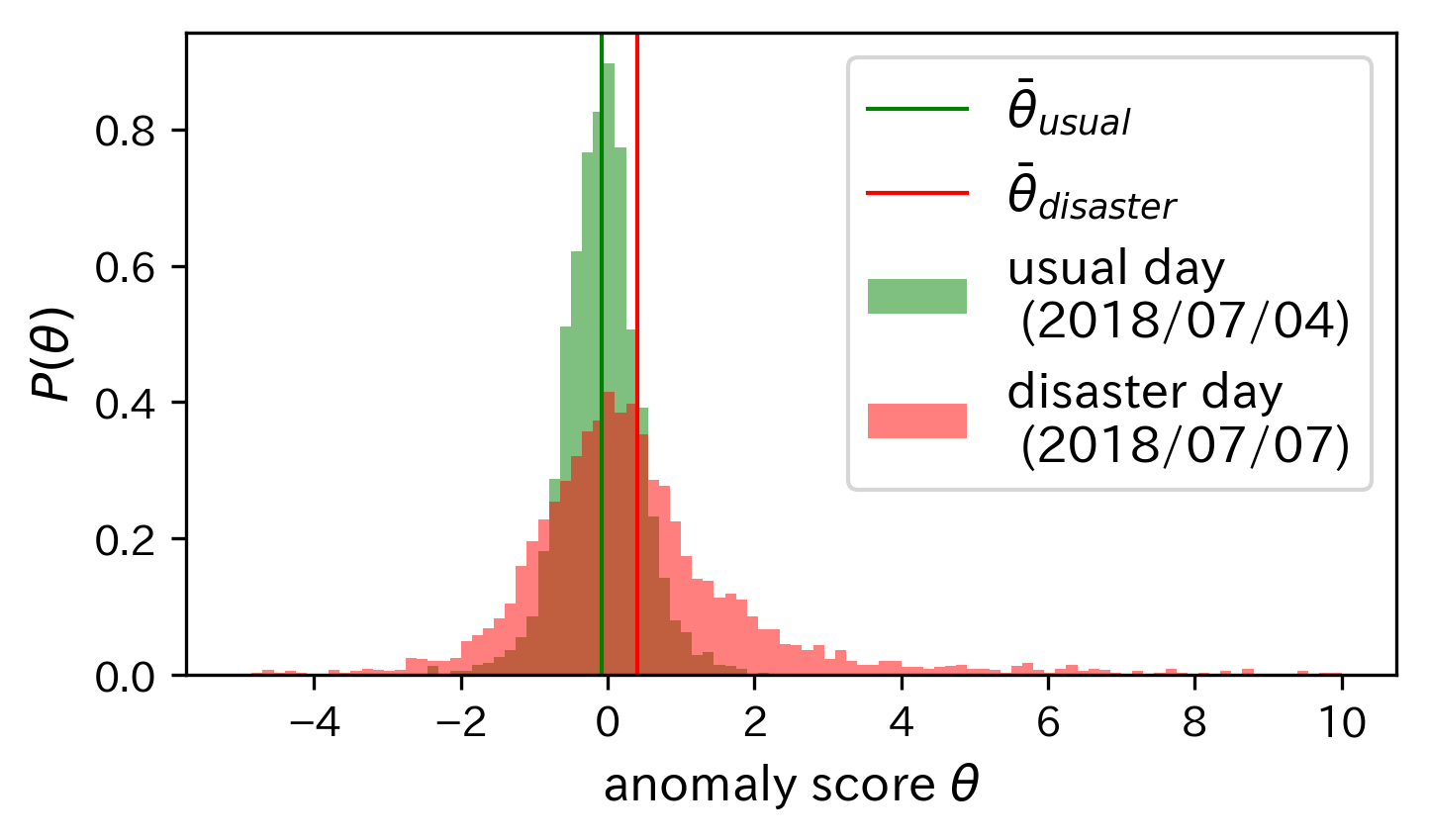}
\caption{Histogram of anomaly scores on a usual day (green) and the disaster day (red). 
% More people have higher anomaly scores on disaster day, indicating evacuation activities.
} 
\label{scores}
\end{figure}

The time parameters defined in Subsection \ref{ad} were set to $t_0 = 2018/6/1$ 12:00AM, $t_l = 2018/7/1$ 12:00AM, $t_d = 2018/7/7$ 1:30AM, and $t_e = 2018/7/10$ 12:00AM. 
We only used web query data collected before the evacuation alert $t_d$ to generate user representations with our \texttt{SQE} model. 
Parameters for the anomaly detection algorithm were set to $\sigma_x = \sigma_y = 100 \text{ (m)}$ and $\sigma_t=1 \text{ (hour)}$. 
$\sigma_x$ and $\sigma_y$ values reflect the spatial error of the location information observations. 
Past works have shown that average spatial errors of location data obtained from smartphones are typically large (several 100 meters) especially when users are located indoor \cite{nishi2014hourly}. 
Also, $\sigma_t$ was set to 1 hour to cover sparsely observed time periods (e.g. nighttime). 
Figure \ref{scores} shows the histogram of anomaly scores $\theta_i$ of individual users. 
We observe higher density with high anomaly scores on disaster day (2018/7/7) compared to a usual day (2018/7/4). 
We chose the value of $\Tilde{\theta}$ using results from a survey\footnote{\url{https://mba.pu-hiroshima.ac.jp/pdf/h30/180901a.pdf} (in Japanese)} carried out by the Hiroshima Business and Management School, which found that only 4.6\% of the affected people evacuated during the 2018 Japan flood. 
We set the threshold to $\Tilde{\theta}=4$ so that the percentage of users with anomaly scores exceeding the threshold is close to the reported evacuation percentage. 
% To label users who are behaving differently their ``usual'' mobility patterns, the threshold value for detecting evacuation activities was set to $\Tilde{\theta}=4$.
The threshold for detecting ``usual'' mobility patterns was set to $\Tilde{\theta}_l = 1$ because anomaly scores are usually within $|\theta| < 1$ on a usual day (Figure \ref{scores}). 
The predictability of evacuation decisions under different parameter values of $\Tilde{\theta}$ are tested in Section \ref{results}. 

Because the problem is a binary classification task ($y_i \in \{0,1\}$), we used prediction accuracy and area under curve (AUC) as metrics for predictive performances. 
% We implement all algorithms using \texttt{scikit-learn} packages.
To prevent overfitting and to ensure that our results are statistically significant, we performed a cross validation with $k=5$ folds.
% , and only report statistically significant accuracy scores with standard deviation less than 5\%.

\subsubsection{Comparative Methods}
% 坪内さん　比較手法の簡単な説明
% \textcolor{red}{Tsubouchi-san}
% We compare the classification performances when using the 
The following methods for generating user representations were tested. \\
\textbf{Input Query Selection.} 
% Since many users have multiple web search queries, it is important how to select the queries to use for generating representations. 
We selected input queries used for feature generation based on 2 criteria. 
First, we tested selecting only a single query as input for \texttt{SSQE} and selecting multiple (max. 10) queries as input for \texttt{SMQE}. 
Second, we tested selecting the most recent query (queries) and selecting high-importance queries defined by \texttt{tf-idf} values. 
% In this case,  \texttt{tf-idf} values were calculated a ``document'' is a bag of query words searched by each user. 
Thus, we tested a total of 4 combinations ([Single, Multiple]$\times$[Recent, \texttt{tf-idf}]) for selecting the set of input queries. 
\\
\textbf{Feature Generation.} 
Using the set of input queries, we generated features using \texttt{SQE} (\texttt{SSQE} or \texttt{SMQE}, depending on the number of input queries). 
To evaluate the predictive performance using the user representations, we compared performances with the case using one-hot encoding of queries (e.g. ``cat''=[1,0,...,0], ``dog''=[0,1,0,...,0]). 
Thus, combined with the 4 input query selection methods, we tested a total of 8 combinations ([Single, Multiple]$\times$[Recent, \texttt{tf-idf}]$\times$[One-Hot, \texttt{SQE}]) for feature generation. 
\\
\textbf{Classifier.} 
Because our main contribution is to show how well user representations generated from web search queries can predict evacuation decisions, the classifier we use to make predictions is not the focus of this paper. 
Thus, we used random forest (RF) which is a standard machine learning model for classification tasks. 
To evaluate the performances of each feature generation method fairly, we used default parameters for RF, with 100 trees and Gini impurity to measure split quality.
For more detail, please see \texttt{scikit-learn}\footnote{\url{https://scikit-learn.org/stable/modules/generated/sklearn.ensemble.RandomForestClassifier.html}}. 

\subsection{Results} \label{results}
% % 坪内さん
% % \textcolor{red}{Tsubouchi-san}
% \subsubsection{Visualizations of Web Search Behavior Representations}
% \textcolor{red}{A nice plot that shows that the representation learning is actually working}

\begin{table}[t]
\caption{Prediction accuracy of evacuation decisions using different methods for generating features from pre-disaster web search data.}
\begin{center}
\begin{tabular}{c|ccc|cc}
\toprule
\multirow{2}{*}{Method} & \multicolumn{2}{c}{Input Queries} & Feature  & \multirow{2}{*}{\textbf{Accuracy}} & \multirow{2}{*}{\textbf{AUC}} \\
& \# words & Metric &  Generation &  &   \\
\midrule 
1 & Single & Recent & One-hot & 0.571 & 0.590 \\
2& Single & Recent & \texttt{SSQE}    & 0.787 & 0.812 \\
\midrule
3& Single & \texttt{tf-idf} & One-hot & 0.733 & 0.600 \\
4& Single & \texttt{tf-idf} & \texttt{SSQE}    & 0.814 & 0.833 \\
\midrule
5& Multiple  & Recent & One-Hot & 0.689 & 0.637 \\
6& Multiple  & Recent & \texttt{SMQE}    & 0.756 & 0.748 \\
\midrule
7& Multiple  & \texttt{tf-idf} & One-Hot & 0.759 & 0.678 \\
8& \textbf{Multiple} & \textbf{\texttt{tf-idf}} &  \textbf{\texttt{SMQE}} & \textbf{0.840} & \textbf{0.837} \\
\bottomrule
\end{tabular}
\label{pred}
\end{center}
\end{table}

\subsubsection{Prediction Accuracy}
% 坪内さん
Table \ref{pred} reports the prediction accuracy of evacuation decisions using different methods for feature generation. 
% Because of the design of the \texttt{SQE} models, 
\texttt{SSQE} (Session-based Single Query Encoder) was used for single query inputs, and \texttt{SMQE} (Session-based Multiple Query Encoder) was used for multiple query inputs.
Out of all 8 combinations of feature generation methods, Method 8, which used \texttt{SMQE} to generate representations of multiple search queries selected by \texttt{tf-idf} scores had the highest prediction accuracy of 84\%. 
The AUC score was also highest for this method, showing the robustness of this result. 
This result can be interpreted as the following: given the web search queries of Yahoo Japan users from prior to the disaster, we are able to predict whether or not that user will evacuate during the disaster with 84\% accuracy.  

Most importantly, we can observe that using \texttt{SMQE} to generate features from input queries significantly improves the predictive accuracy over using One-Hot vectors ($\times$110.7\% in accuracy and $\times$123.4\% in AUC).
This is mainly due to the characteristic of query representations that can capture subtle differences between the queries, as discussed in more detail in subsection \ref{analrep}. 
Prediction scores of \texttt{SMQE} were higher than \texttt{SSQE} in all cases (Method 2 vs 6, 4 vs 8), which implies that increasing the number of query words used to generate representations also improved the accuracy of evacuation prediction. 
It was counter-intuitive to observe that rather than using queries searched by the users just before the disaster (``recent''), using queries with high \texttt{tf-idf} values had higher predictability when other settings were identical (Method 1 vs 3, 2 vs 4, 5 vs 7, 6 vs 8). 
% Thus, in most of the cases users search topics related to evacuation some time before the evacuation timing. 
Intuitively, we may hypothesize that users tend to search about things related to evacuation more as the situation becomes worse (and gets closer to evacuation timing).  
However, this result implies that our intuition is incorrect, and that selecting the input queries based on importance (e.g. \texttt{tf-idf}) enables higher predictability.

\begin{figure}[t]
\centering
\includegraphics[width=0.5\columnwidth]{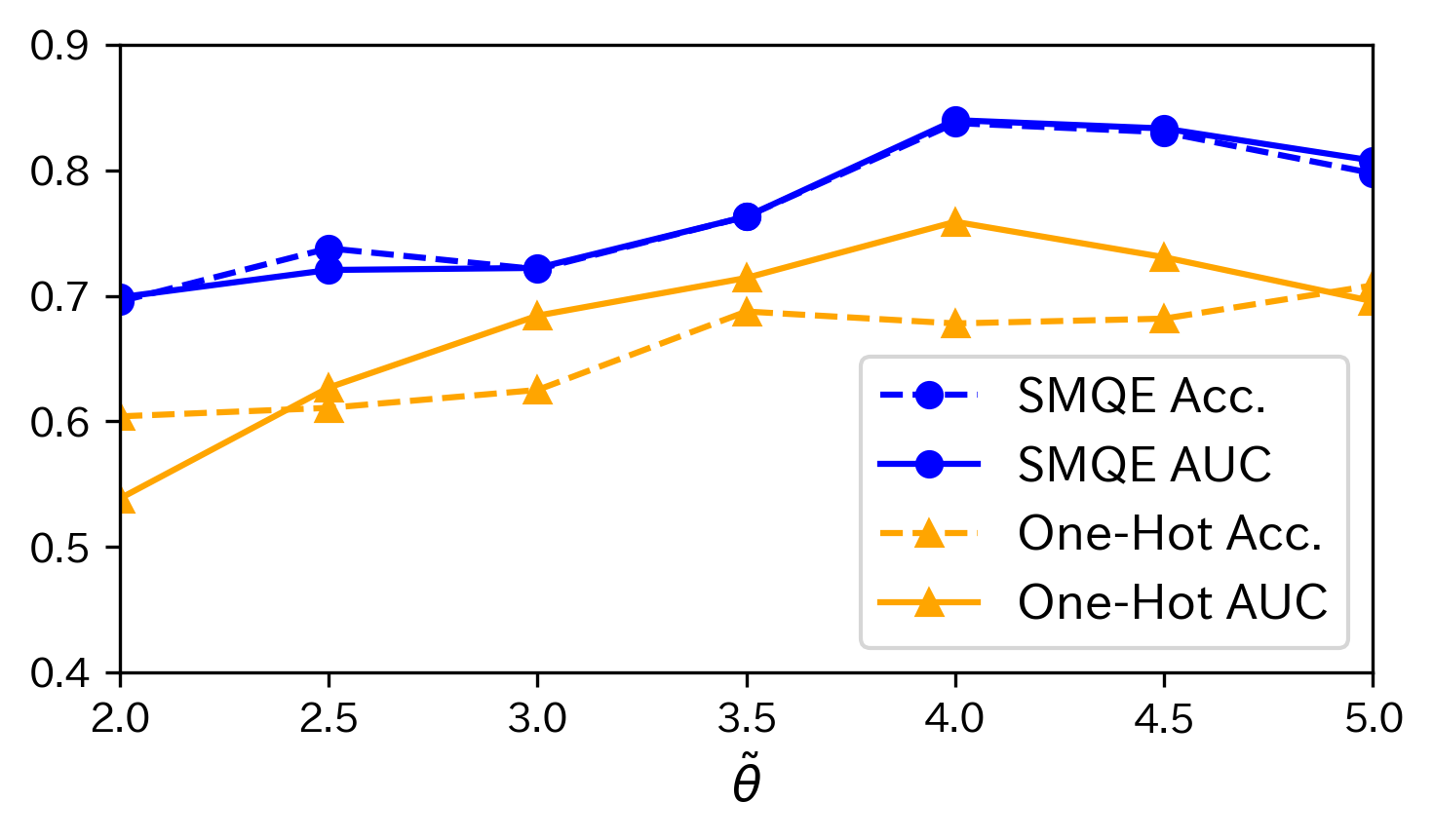}
\caption{Sensitivity of prediction scores (accuracy, AUC) with respect to different values for $\Tilde{\theta}$. } 
\label{acc}
\end{figure}

Figure \ref{acc} shows the prediction scores of Methods 7 and 8 under various values of the threshold parameter $\Tilde{\theta}$. 
For all  $\Tilde{\theta}$ values, Method 8 had higher accuracy and AUC compared to Method 7, clarifying the significant improvement of using \texttt{SMQE} over One-Hot encoding. 
With smaller $\Tilde{\theta}$, the confidence of the anomaly detection algorithm would decrease, meaning that we are less certain on whether the user did or did not evacuate. 
Thus, the general trend that the prediction accuracy decreases as we decrease  $\Tilde{\theta}$ was expected.
However, the prediction accuracy of Method 8 stayed high (AUC = 0.7) even when  $\Tilde{\theta}=2.0$, while the AUC of Method 7 dropped to 0.53. 
This shows that our method is capable of classifying users even when the observed post-disaster mobility patterns were unclear, due to sparse and/or noisy observations. 

% \textcolor{red}{Figure showing how prediction accuracy changes when we change threshold $\Tilde{\theta}$}

\subsubsection{Analysis of Representations}
\label{analrep}
% \textcolor{red}{Top words that appear among users who were predicted correctly (for both sides)}% 坪内さん
% Table \ref{words}
In addition to showing the high predictability of our proposed method, looking into how the models treated the input queries enables us to understand how the high predictability was achieved. 
% Table \ref{words} shows some examples of queries that were correctly classified using model 8 but were incorrectly classified using model 7. 
When we compare prediction results between Method 7 and Method 8, queries such as ``manga cafe'' (a cafe-like space where you can read comics and also stay the night) and ``all-night restaurant'' (restaurants where you can stay the night) were correctly used to predict evacuation decisions using Method 8, because the \texttt{SMQE} model generated representations close to evacuation shelters and hotels. 
In addition, because of the accurate representations, Method 8 was able to prevent false predictions compared to Method 7. 
For example, Method 8 was able to classify specific (local) and sparsely searched location names (``Suna-gawa'', ``Kumano-cho'', ``Tamashima'') correctly compared to Method 7.
Method 7 could not classify such kind of subtle variances of location names, and was limited to encoding names of large districts that were searched frequently (e.g. ``Kurashiki City'', ``Okayama prefecture''). 
These examples show how \texttt{SMQE} was able to generate representations of queries so that subtle differences in vocabulary and names of places were captured, which was not possible using one-hot vector representations. 

% \begin{table}[t]
% \caption{Examples of Queries correctly classified by Model 8 but not by Model 7}
% \begin{center}
% \begin{tabular}{l|l}
% \toprule
% \textbf{Queries} & \textbf{Description} \\
% \midrule
% ``Manga-cafe'' & \multirow{2}{*}{POIs with hotel-like functions} \\
% ``All-night restaurant'' \\ 
% \midrule
% ``Suna-gawa''  &  \multirow{3}{*}{Specific names of locations}  \\
% ``Kumano-cho'' &   \\
% ``Tamashima'' & \\
% \bottomrule
% \end{tabular}
% \label{words}
% \end{center}
% \end{table}

\section{Discussion}
Our experimental results using real world data confirmed that evacuation decisions are highly predictable by utilizing the users' web search behavior observed prior to the disaster. 
This method proposes an alternative method to predicting evacuation behavior, which overcomes the drawbacks of depending on sensitive real time location data. 

Now, we discuss future research opportunities that this study enables. 
First, the data we could use to test and validate our method was limited to one disaster event (2018 Japan floods).
Once we start collecting additional data from future disasters, testing the generalizability of our method between different disaster events (future floods) will be our focus of future research. 
Moreover, testing whether a model pre-trained using data from a given disaster would work on other types of disasters would be an ambitious but very interesting research question to investigate. 
Even with our limitation on data in this study, we were excited to see that web search behavior can indeed predict evacuation decisions.  

Second, in this study, we did not use mobility data as inputs for prediction. 
This was because we wanted to investigate the predictability using only the web search behaviors, and also assumed the situation where we are not able to use location data at all due to privacy issues. 
As shown in previous studies, using pre-disaster (usual) mobility patterns is known to improve the accuracy of post-disaster behavior \cite{yabe2016predicting}. 
We are also interested in building a framework that can integrate both pre-disaster physical and web search behavior for post-disaster evacuation prediction. 

Thirdly, we found that evacuation decisions can be predicted with high accuracy using just the web search behavior. 
This motivates us to predict evacuation destinations from web search behavior as well, although it is known that most of the evacuees head to shelters or familiar locations such as houses of family and friends \cite{lu2012predictability}. 
This exciting extension to our current work would require prior knowledge on where the users typically visit using location data, thus would be an additional topic to our second point.

Finally, we touch upon the industrial applications that will be developed as a product of this research.
In contrast to most existing studies, evacuation decisions were predicted based on web search behavior representations generated from data observed \textit{before} the evacuation alert, using web search data which have less concerns on privacy issues compared to location information. 
This method is feasible in real world settings, by collecting web search data and making predictions in real time, before the disaster alerts are sent out.  
Using this output, we could support decision making of local governments by providing predictions of how many people will evacuate prior to the disaster. 
This information could help policy makers to organize their strategies on the allocation of shelters and emergency supplies.  
Moreover, towards the individual users, we could provide informational support, such as notifications to their smartphones about information of nearby evacuation shelters, road closures, and dangerous areas to avoid. 

\section{Related Works}
\subsection{Human Mobility Analysis during Disasters}
% 矢部
Traditionally, surveys have been used to understand evacuation decisions after disasters \cite{sadri2018role,sadri2017modeling,mesa2012household}.
% Such surveys have been utilized after disasters as well, and have shown to be useful in understanding the actions of evacuees in detail \cite{sadri2018role}. 
Recently, large scale datasets 
% collected from mobile phones 
are being utilized to understand the behavior of individuals 
% with low cost 
\cite{gonzalez2008understanding,calabrese2011estimating,yuan2012discovering}.
% Many works have applied this new data source for disaster management  applications \cite{gething2011can}.
% Studies have shown the effect of weather patterns on human mobility \cite{horanont2013weather}, and its predictability using socio-economic factors \cite{yabe2016predicting}.
% To understand human mobility during larger scale disasters, 
Lu et al. analyzed call detail records to investigate the predictability of evacuation destinations 
% of individuals 
after the Haiti earthquake \cite{lu2012predictability}.
% Wang et al. used Twitter Geo-tag data to show that the mobility patterns of the victims of Hurricane Sandy were perturbed significantly \cite{wang2014quantifying}. 
Song et al. revealed the population decline in various areas after the Great East Japan Earthquake using GPS data \cite{song2013intelligent}. 
More recent studies showed that evacuation behavior could be monitored after an earthquake by using mobile phone location data collected from evacuees \cite{yabe2016estimating,wilson2016rapid,yabeeq}. 

In addition to the literature on post-disaster analysis of evacuation behavior using big data sources, there have been various works that have attempted to predict mobility patterns using real time information.
Fan et al. and Sudo et al. use real time location data to predict human mobility dynamics in an online manner \cite{sudo2016particle,fan2015citymomentum}. 
Other works using machine learning models have been proposed for real time prediction \cite{song2013modeling,song2014prediction,song2017prediction}.
More recently, Jiang et al. proposed \texttt{DeepUrbanMomentum}, a deep-learning architecture that models the human mobility dynamics \cite{jiang2018deepurbanmomentum}. 
Although these frameworks have been shown to be effective in predicting short term future mobility patterns, they require real time location information, which are becoming harder to obtain due to rising privacy concerns. 
Thus, we take an alternative approach and attempt to utilize other data sources (web search behavior) to predict evacuation mobility patterns. 
A closely related work was performed by Konishi et al., which used transit app queries to predict future destinations of people \cite{konishi2016cityprophet}. 
In their study, input data were names of stations in the transit system. 
Our study applies a similar idea but with web query data, which do not directly reveal the users' destinations. 
Web queries contain latent meanings and subtle differences in expressions, which makes our prediction task much more harder.

\subsection{Representation Learning of Web Search Behavior}
% 清水さん、お願いいたします。見出しも適切なものに変更して下さい。
% \textcolor{blue}{Shimizu-san} \\
The main component of this work is to understand users' web search behavior using a model that can encode sessions of search queries into representations that reflect their latent meanings and underlying search intent of the user. 
There have been many studies on such encoders for words and texts.
Some of the popular approaches are word embedding methods such as \texttt{word2vec} \cite{mikolov2013distributed}, \texttt{FastText} \cite{joulin2016fasttext}, and \texttt{GloVe} \cite{Pennington14glove:global}.
While these methods provide relatively low-cost, easy-to-use ways to generate word vectors, it is not clear how to combine multiple word representations, for example embedding phrases.

Also, since these methods directly assign vectors to a word (or possibly a query itself when the target domain is search queries), it is difficult to handle the tail part of the query sequence that contains too many words. 
In our method, we avoid this difficulty by applying LSTM models to the character sequence and by generating the representation compositionally. 
One of the more closer attempts to develop such encoder is \texttt{Query2Vec} \cite{kang2015query2vec}.
While it has a similarity with our method in the usage of the search query data and the user's web search behavior, our method handles the query's text representation differently.
In \texttt{Query2Vec}, a dense vector is assigned to each word or query, similar to the word embedding models, thus suffers from the same aforementioned drawbacks.
Also, while \texttt{Query2Vec} can produce representations of words and queries, it is unable to produce representations of sessions of queries.

% Also, a nature of these method that a vector is directly assigned to a word (or possibly a query itself when the target domain is search queries) make it difficult to handle the tail part of the query distribution which contains too many words to manage as the vocabulary and remember their corresponding vectors.
% We avoided this difficulty by applying LSTM models to the character sequence and generating the representation compositionally. 
% One of more closer attempts to develop such encoders is \texttt{Query2Vec} \cite{kang2015query2vec}.
% While it has a similarity with our method in the usage of the search log and user's behavior, there is a difference in how to handle the query's text representation.
% In \texttt{Query2Vec}, a dense vector is assigned to each word or query, so it carries the same difficulty with word embedding methods.
% Also, while \texttt{Query2Vec} can produce representations of words and queries, it is unable to produce representations of sessions of queries.

\section{Conclusion}
% Disasters impose a great threat on the lives of individuals, and recent events have shown the importance of effective first response and relief efforts. 
Despite the importance of predicting evacuation activities, most works have used real time location information, which are becoming increasingly difficult to obtain in reality. 
In this paper, we bridge this gap by proposing a framework that utilizes web search queries of users to predict post-disaster evacuation decisions. 
Through experiments using real world data, it was found that evacuation decisions are highly predictable by learning the user representations from web search queries. 
This study opens up a new avenue of research on the prediction of mobility using web query data, and encourages further studies using various similar datasets. 
Moreover, this study proposes an alternative method for predicting evacuation prior to the disaster, which has huge potential for applications in real world disaster situations. 

\section*{Acknowledgements}
T.Y. is partly supported by the Doctoral Fellowship provided by the Purdue Systems Collaboratory. The work of T.Y. and S. V. U. is partly funded by NSF Grant No. 1638311 CRISP Type 2/Collaborative Research: Critical Transitions in the Resilience and Recovery of Interdependent Social and Physical Networks.

\bibliographystyle{unsrt}  
\bibliography{references}  %%% Remove comment to use the external .bib file (using bibtex).

\end{document}